# Matter-antimatter asymmetry and chiral behavior in perturbed atom H.

G. Van Hooydonk, Ghent University, Faculty of Sciences, Krijgslaan 281 S30, B-9000 Ghent (Belgium)

(e-mail: guido.vanhooydonk@rug.ac.be)

Recently, we showed that band, as well as line spectra reveal a left-right symmetry in 4- and 2-fermion systems. We now show how a *mass-conjugation* explains the difference between atom- and antiatom states of a stable 2-unit charge Coulomb system. For the electron-proton system in natural H, this difference (*an anomaly*) between states belonging to the *hydrogen (matter) world* and the *antihydrogen (antimatter) anti-world* is governed by number $2m_e/m_H = 0.0011$. This is verified with the H line spectrum. Classical physics can describe the parity effects for matter-antimatter *asymmetry*. An *internal* mass-conjugation (*internal* algebra) extends and generalizes Dirac's fermion-anti-fermion *symmetry* and defies Einstein's Special Theory of Relativity (STR).

Pacs:

## 1. Introduction

One can never overestimate the importance of the H *line spectrum* to understand the laws of nature. Atomic 2-fermion system H played a decisive role in the development of quantum theory and is important for metrology [1]. Today, many years after Bohr theory, only bound state QED [2], based on Dirac-Einstein theories in a Coulomb field, can account for this *line spectrum* with great precision (errors $< 10^{-11}$). Similarly, molecule $H_2$ and its *band spectrum is the only benchmark* available for neutral, stable 4-fermion Coulomb systems. Today, we understand the $H_2$ *band spectrum* only with Schrödinger wave-mechanics-based Heitler-London theory, published in 1927 [3].

Despite the achievements of these 2 complex theoretical frameworks for benchmark systems H and $H_2$, problems persist with the *chiral behavior of neutral particles.* Matter-antimatter (a)symmetry, a mirror or left-right (a)symmetry (a parity) and its implicit mass-conjugation are important for unification. *Particle-antiparticle symmetry for charge-conjugated particles (fermions) goes back to Dirac theory.* Although with charged particles, charge conjugation is sufficient to arrive at mirror symmetry (parity, inversion), the problem is more difficult when neutral composite particles are involved, since mass-conjugation is not evident between 2 neutral positive mass particles. There is a *Babel of Tongues* for matter-antimatter (a)symmetry [4]. This is the broader context of this report: the morphological (bio-) diversity in nature suggests chiral behavior is a generic property of matter.

Using the molecular band spectrum of $H_2$, we recently found [5] that *atom-antiatom symmetry* really exists. To make sense, a single atom must show this very same symmetry: a natural atom H must be able to take the morphology of a classical *atomic-* as well as that of an *anti-atomic*-state. The unavoidable consequence -the challenge- of [5] was to detect atom- and antiatom-states by means of an atomic line spectrum. We showed [6,7,9] that such a spectrum points towards *intra-atomic* left-right or *mirror symmetry* also, exactly as we anticipated [5]. *Using QED one-electron energies for Lyman ns-singlet states* [8], *algebraic effects on a left-right transition in the neutral, bound and stable 2-unit charge Coulomb system in H are easily exposed quantitatively* [6,7,9]. These unprecedented results [5-7,9] are not directly visible in the 2 complex theoretical frameworks available: bound state QED [2] for atom H and Heitler-London theory [3] for molecule $H_2$.

In addition, using conventional Heitler-London theory it is still difficult to assess the stability of a hydrogen-antihydrogen or H$\underline{H}$ complex [10-16]. One suspects (*atom-antiatom*) *annihilation* but it remains difficult to tell exactly how stable 4-fermion H$\underline{H}$ is, if it were stable at all. This is surprising, since similar 4-fermion HH system in $H_2$ and its *band spectrum* can be described quite exactly [17,18].

Recent results [5-7,9] defy 2 standard theories. So, *mirror symmetry* at the neutral particle level must be reassessed. Normally, we take most of *its* implications (inversion in 1D) for granted or consider them as trivial, since a right-handed reference frame in front of a mirror (plane) is transformed into a left-handed one behind the mirror (plane). As detailed in [9], much depends on the position of a mirror (plane): is it external or internal with respect to left- and right-handed objects? Is an observer situated *in- or outside* the system exhibiting chiral behavior [9]? The mirror image of a *real object* in front of a *real mirror* is an exact but differently handed copy of that object. Seen in the mirror, the object is *imaginary or virtual* and *cannot* be observed in the world of external observers. In this case, a *real mirror* is placed outside a *real object*, whatever its absolute handedness, which is inverted in the virtual object seen in (or *behind*) a mirror (plane). *This case suggests mirror symmetry is a discrete and virtual symmetry.*

For *two real objects*, left-/right-handed, say left and right hands, *it only appears there is a mirror in between. In reality, there is no mirror, since the details of a left hand are different than those of a right hand.* With 2 real objects, differently handed, the *mirror* in between does not really exist since there is *left-right* *asymmetry*. This difference in the details of real left- and right-handed objects can never be reflected in a real mirror[1], which is at the roots of *chiral behavior* as observed in nature [9]. *The picture gets more explicit realizing that also a hand's front and back are differently handed.* The differences, the *asymmetry*, between front and back of a hand are much larger than those between the fronts or backs of a left and a right hand, indicating that there are *gradations in chiral behavior* or *that mirror symmetry is not just a discrete and virtual symmetry, since it can also be continuous and real.* There is no real mirror plane *within a hand*, despite the left-right topology of front and back, *since chiral behavior or a left-right asymmetry has taken over*. Mathematical, physical and chemical effects of chiral behavior are given in [9].

Trying to measure a difference between atom and anti-atom energies is a goal of current CERN/Fermilab *antihydrogen* experiments [19]. For the CPT-theorem, the value of the 1S-2S-transition in H̲ is critical. But, if left-right (a)symmetry is fundamental for neutral particle systems observed in nature and if neutral species H is their benchmark, there must be a connection between the H spectrum and a mirror (a)symmetry. Taking the above hand metaphor, we expect that *back and front versions of atom H* -whatever these stand for- are different. This is exactly what we already found: an *intra-atomic mirror symmetry* [5-7,9]. *But in the advent of unification, this mirror (a)symmetry must ultimately be understood with algebraic mass effects (mass-conjugation) with 2 masses on a single axis (the field).*

In [5-7,9] we used a phenomenological approach to mirror symmetry at the neutral particle level, the atom. In this report, we try to expose the rationale behind these phenomena and to focus on the existence of and the transition between 2 mass-conjugated worlds, *a matter and an anti-matter world*.

To get at a chiral field effect for the electron-proton interaction in H, *we must find a mass-conjugation, whatever its nature, in a single world*, i.e. the real world where all of our observations are made.

## 2. Matter-antimatter asymmetry

The 2 mass-conjugated worlds are: a *matter-world*, were particles (neutral or charged, fermions or bosons) have positive mass and an *antimatter-world*, where these particles have negative mass. This is just a uni-axial 1D representation of matter-antimatter symmetry like for algebraic numbers on an axis [9]. As in previous reports [6,7,9], we use standard Bohr $1/n^2$ theory as a guideline for detecting matter-antimatter (a)symmetry at the atomic level, not bound state QED, where fermion spin (helicity) accounts for chiral effects [2]. A point-like particle with positive mass (matter, right...) in front of a real mirror (an inversion point, an origin of an axis) is *inverted* into a *virtual* point-like antiparticle with negative mass (antimatter, left...) [9]. We can use a parity operator **P** for particle mass m, giving eventually $(-1)^\mathbf{P}$m or +m for **P**=0 and –m for **P**=1, exactly the parity operator for algebraic numbers on an axis.

---

[1] A mirrored left hand is a *virtual right-handed left hand, with all physical details of a left, not those of a right hand.*

To make the approach generic, we use a *zero mass perturbation* of hydrogen mass, as H is the prototype of a bound 2-unit charge system [9]. In this 1D view, *static* atom H in a positive matter world has energy

$E(+)/c^2 = +m_H$

Anti-hydrogen atom H̲ in a negative antimatter world has energy

$E(-)/c^2 = -m_{\underline{H}}$

Mass $|m_H|$ is equal in both worlds and c is the velocity of light.

*Which world is actually designed as positive or negative and which of the two is the real world, depends on convention. The (mass-) exclusion principle is that a negative (positive) mass particle is forbidden in the positive (negative) mass world.* We correlate signs of mass and energy, although, conventionally, *mass and energy are conjugated* (positive mass atoms have negative total energy).

When *a zero mass perturbation* is applied to atom H, mass $m_H$ is left unaltered, whatever its sign. Despite its simplicity (the justification is given below), this model leads to pertinent results, significant to get a matter-antimatter (a)symmetry and a mass-conjugation.

The model with $\pm m_H$ is a substitute for *a symmetrical achiral division* (bisection) of mass $m_H$ in 2 halves $+½m_H$ and $-½m_H$, centered at a local origin (the scale factor is 1 for 2 species and ½ for 1 species only), *a description for which a negative world is allowed*. A *mathematical* metaphor for a zero mass perturbation is that a positive number +1000 can be described as an average of 2 positive numbers like +999 and +1001. We use a local origin at +1000 and 2 auxiliary numbers +1 and –1 (*the known measure or unit for numbers*) at either side of this local origin [9]. Without perturbation, the alternative for achiral division (*bisection*) is to describe positive number +1000 as a line segment with vertices – 500 and +500 centered at a common local zero origin (*but situated at +500 in the absolute reference* *frame*). The analogy between the algebra of pure numbers and that of *physical* numbers (i.e. scaled zero-dimensional physical parameters like $m/m_0$), including the case of *complex* numbers, is dealt with in detail in [9].

The *physical justification* for using a zero mass perturbation is simple. We can only study H by perturbing it with heat (kT), radiation (hν).... Its effect can be estimated by comparing energies in unperturbed and perturbed systems. Perturbed H has energy: $[E_H+kT]$, $[E_H+h\nu]$ or $[E_H+hc/\lambda]$. Using $E_H=m_Hc^2$, we have $E_H(pert) = E_H[1+ kT/(m_Hc^2)]$. Reduced perturbation $kT/(m_Hc^2)$ is *an extremely small number*, even at high temperatures. With twice the Rydberg or reduced mass as alternative unit energy $E'_H$ for H ($\mu\alpha^2c^2$ instead of $m_Hc^2$), this ratio is $10^8$ times larger (~ $10^{-5}T/°$). As a generalization, the algebraic effect of *a very small perturbation, a generic zero mass perturbation*, on a system like H with mass $m_H$ is of crucial importance. How to scale the corresponding energies is another problem. With reduced mass for instance, the matter-antimatter worlds are rescaled to give

$$E/(\alpha c)^2(\pm)=\pm\mu$$

instead of absolute $E/c^2(\pm)=\pm m_H$. A *reduced mass scale* is the basis of the present analysis (see also [9]). Scale factor $\alpha^2$ is not relevant for the algebra (symmetry) of the worlds.

The example with numbers +999, +1000 and +1001 illustrates the possibility that gradations in chiral behavior are also plausible. *Standard unit difference +1 between numbers N and N+1 becomes less visible with increasing N. Their ratio 1/(1+1/N) is 1 only for infinite N.*

The world-antiworld unitary symmetry (or the algebra of numbers [9]) is perfectly understood with mathematics. The problem of mirror symmetry for *the non-algebraic, exclusively positive numbers in physics* (of type p/q, where p and q are measured quantities having the same dimension) is to verify if mirrored systems or left-right transitions in nature/physics are virtual (forbidden) or real (allowed).

*2.1. Generic zero mass perturbation of atom H in the positive or matter world*

Here, zero mass perturbation

$$+m_H=+m_H\pm 0=m_H\pm(m_{x1} - m_{x2}) \qquad (1)$$

leaves mass $+m_H$ unaffected, if and only if $|m_{x1}|=|m_{x2}|$. Equation (1) is a means to look differently at positive mass atom H without changing its most fundamental characteristic, mass

+$m_H$. The components X of the perturbing mass-conjugated pair have yet unspecified mass $m_{xn}$, probably obeying

$$m_H \neq m_{xn} \qquad (2a)$$

although (2a) is not an absolute condition. Since we are studying H, let electron mass $m_e$ be the unit mass for perturbing system X or

$$|m_{Xn}| = p|m_e| \qquad (2b)$$

where p is an unknown *number* (integer or not, large or small, variable or not). The value of p does not affect the algebraic mass symmetry in the perturbing pair.

*A generic zero mass perturbation of one (neutral) particle with positive mass as in (1) always gives 3 particles instead of one, of which 1 must have negative mass, which is, by definition, forbidden in a positive mass world. At generic level (1) of a possible chiral theory, there is no need to introduce fields, dynamics or any other physics: only algebraic number theory (real or complex, see below and* [9]*) is involved.* The conflict with negative mass particle $-m_x$ in a positive mass world in (1) can be solved in 2 ways: (a) a conventional one and (b), a solution which defies convention.

*(a) Classical or conventional Bohr solution in a +world*

When Bohr said that, in the *conventional* positive matter-world, atom H consists of an electron with positive mass $m_e$ and a proton with positive mass $M_p$, he actually proposed to interpret algebraic mass equation (1), *without speaking of charges*, literally as follows

$$+m_H = +m_H \pm 0 = +m_H - (m_{x1} - m_{x2}) = +(m_H - m_{x1}) + m_{x2} = +(m_H - m_e) + m_e = +M_p + m_e \qquad (3)$$

with $|m_e| < |M_p|$ and $m_{xn} = m_e$ (p=1). With the number metaphor, Bohr here said that positive number $+1000 \equiv +1000 - 1 + 1 \equiv +999 + 1$. The appearance of a negative mass particle $-m_e$ in a positive mass world is avoided, as it should. *Unit $-1$ cannot belong to a positive number world.* Bohr let particle mass $-m_e$ be *absorbed* in a heavier one with positive mass $m_H - m_e = M_p$: only the *internal* sign of the composite particle is negative in the positive mass world, its *external* sign remains positive as $M_p > m_e$. Up to this level, the analogy with algebraic number theory (Dirac's particle-antiparticle symmetry) is evident, if it were not for the algebra *within* the composite baryon, the proton. *The notion of conjugated algebraic charges for the remaining 2 particles is not even necessary.*

However, *classical physics* is involved for describing the interaction between the 2 remaining particles with mass $M_p$ and $m_e$. *To get at a soluble pseudo 1-particle system*, Bohr had to proceed classically with a reduced mass for the 2-particle combination in (3). With (1) as our basis for a chiral field theory and *without having to consider particle charges at this instance*, we get a reduced mass with 2 sign characteristics instead of only one:

(i) the first is *external*: + for the world (the classical *Dirac mirror symmetry*)

(ii) the second is *internal*: – for the *composite* particle, needed to arrive at its positive mass.

As stated above, also reduced mass in a +world must be positive but must be described with external E and internal I algebra (E,I), giving $\mu(E,I)$. Using (3), the result is

$$\mu_1(+,-) = m_e(m_H - m_e)/m_H = m_e(1 - m_e/m_H) = \mu_{Bohr} = m_e M_p/(M_p + m_e) = m_e/(1 + m_e/M_p) \quad (4)$$

*indistinguishable from the original Bohr result with $m_e$ and $M_p$ directly.*

In the present context, we obtain nevertheless *an algebraic reformulation of Bohr's reduced mass* $\mu_1(+,-) = \mu_{Bohr}$.

*(b) Non-conventional mass distribution in a +world.*

Using (1), an alternative distribution of the 3 particle masses in (1) is possible. Instead of (3), we obtain the mathematically perfectly allowed *mass redistribution*

$$+m_H = +m_H \pm 0 = +m_H + (m_{x1} - m_{x2}) = +(m_H + m_{x1}) - m_{x2} = +(m_H + m_e) - m_e \quad (5)$$

*for the same zero mass perturbation*. This distribution is forbidden in the positive mass world *since one of the 2 remaining particles of interest retains its negative mass*. With the number metaphor, we say that $+1000 \equiv +1000 + 1 - 1 \equiv +1001 - 1$. If *classical physics* were involved here too, we can calculate a reduced mass for this forbidden system using the same procedure as above. We obtain

$$\mu_2(+,+) = -m_e(m_H + m_e)/m_H = -m_e(1 + m_e/m_H) = -m_e(1 + 2m_e/M_p)/(1 + m_e/M_p) \quad (6)$$

still without speaking about particle charges.

*This second interpretation (5) of a zero mass perturbation for $m_H$ is forbidden in the positive mass world, since reduced mass (6) is negative. It can only be allowed in the negative mass or anti-matter world.* Forbidden reduced mass (6) correlates with allowed one (4) as

$$\mu_2(+,+)/\mu_1(+,-) = -(1 + 2m_e/M_p) \quad (7a)$$

These two reduced masses do not only have a different external sign (*are externally conjugated* in the Dirac sense) *but there is also a (small) difference in magnitude, due to a different internal sign.* The *anomaly* introduced by means of an algebraic switch in a zero mass perturbation $m_H$ (1) is $2m_e/M_p$ (7a) for the ratio and $2m_e/m_H$ for the difference. Let us rewrite (7a) in terms of hydrogen mass. With $M_p=m_H-m_e$, we obtain

$$1+2m_e/M_p = (m_H+m_e)/(m_H-m_e)=(1+m_e/m_H)/(1-m_e/m_H) \qquad (7b)$$

which is *asymmetrical* in classical recoil $m_e/M_p$ but *algebraically symmetrical* in a hydrogen mass based version of classical recoil $m_e/m_H$ [9b]. Internal mass remains conjugated, an extension absent in Dirac particle-antiparticle theory based on external algebra only. *Classical external mirror symmetry is associated with a parity operator (conjugation) for algebraic numbers* [9].

The number metaphor is illustrative: *a physical approach to +1000 says that, in a +world, approximation +999+1 is allowed but approximation +1001-1 is forbidden, according to the world-antiworld distinction.*

We must now verify what happens if a zero mass perturbation were applied to $-m_H$.

*2.2. Generic zero mass perturbation of antiatom H̲ in a negative anti-matter world*

Zero mass perturbation of *antihydrogen* H̲ with mass $-m_H$ must be written as

$$-m_H=-m_H\pm 0=-m_H\pm(m_{x1}-m_{x2}) \qquad (8)$$

which differs from (1) only by the external sign of mass $m_H$, a trivial result. The internal sign of the perturbing mass-system is unaffected. Using the metaphor above, we discuss number $-1000$ for which we have symmetrical descriptions $-999-1$ and $-1001+1$, where only $-1$ is the (*known*) unit for negative numbers.

Using (2) and the same conventions as in Section 2.1, it is easily verified, without giving details, that 2 reduced masses are also obtained in the antiworld. The third in the series is

$$\mu_3(-,-)=(-m_e)[-(m_H-m_e)]/(-m_H)=(-m_e)[-(m_H-m_e)]/(-m_H)=-m_e(1-m_e/m_H)=-m_e/(1+m_e/M_p) \qquad (9)$$

equal to $-\mu_1(+,-)$ or (4). *This is simply the negative of the Bohr reduced mass, allowed in the negative world.*

Finally, the fourth reduced mass is

$$\mu_4(-,+)=m_e[-(m_H+m_e)]/(-m_H)=m_e(m_H+m_e)/m_H=m_e(1+m_e/m_H)=m_e(1+2m_e/M_p)/(1+m_e/M_p) \qquad (10)$$

equal to $-\mu_2(+,+)$ or (6). As in (7a), we obtain

$$\mu_4(-,+)/\mu_3(-,-) = -(1+2m_e/M_p) \qquad (11)$$

identical with (7a) and reproducing result (7b).

*Third reduced mass (9) is allowed in the anti-matter world and forbidden in the matter world. Fourth (10) is forbidden in the anti-matter world but is allowed in the matter world.* This is a mathematical consequence of trying to describe a neutral boson ($m_H$) as a neutral 2-fermion system as Bohr did in his model for atom H. The result from this analysis is that in a positive matter world (the Bohr world for H), mass $m_H$ is forced to get an external algebraic sign, which finally reflects the internal symmetry of the 2-fermion Coulomb system with 2-unit charges, which we did not yet consider. Due to a world-antiworld conjugation, 2 reduced masses are nevertheless allowed: $\mu_1(+,-)$ in (4) and $\mu_4(-,+)$ in (10), *the latter being neglected by Bohr.*

What about the introduction of particle charges? Obviously, these two different external signs for reduced masses reflect the signs of a Coulomb power law, confined to one world only. *The important transition from a reduced mass model without* algebraic particle charges *to a Coulomb model with* algebraic particle charges *is simple classical physics, if total particle energy is introduced (see Section 3.2). Whatever sign we choose for the Coulomb- or charge-based-world (a singlet), the reduced mass for the 2 particle system in the mass-based world corresponding with this Coulomb-world will always be a doublet, separated by twice recoil $2m_e/M_p$.*

Although algebraic switches for masses *are simply inverted, the quantitative results for reduced masses in a zero mass perturbation scheme are indeed different since*

$$\mu_4(-,+)/\mu_1(+,-) = (1+2m_e/M_p) = (1+m_e/m_H)/(1-m_e/m_H) \qquad (12)$$

as in (7b). Quantitatively, *the anomaly of the 2 allowed reduced masses in the matter world is*

$$\mu_4(-,+)/\mu_1(+,-) - 1 = 2m_e/M_p = 0.0011\ldots \qquad (13)$$

These subtle algebraic mass effects are not covered by Bohr's use of the Coulomb law. Should we now refer to bound state QED, *where chiral behavior is connected with fermion spin*, to solve these new simple results or should we proceed as classically as possible with Bohr theory? In fact, it may be possible to retrace the effect of *algebraic mass asymmetry* connected with the world-anti-world transition for a bound 2-unit charge Coulomb system directly *by considering 2 different reduced masses*

$$\mu_1 = -\mu_3 = m_e(1-m_e/m_H) = \mu_{Bohr}$$

$$\mu_4 = -\mu_2 = m_e(1+m_e/m_H) \qquad (14)$$

leading to a reduced difference

$$(\mu_4 - \mu_1)/m_e = 2m_e/m_H = 0.0011\ldots \qquad (15)$$

*If these derivations are valid, ratio $2m_e/m_H$ may be a fundamental constant of nature, related finally to the matter-antimatter asymmetry in the bound 2-unit charge Coulomb system in H* [9]. Using the above metaphor, number $2m_e/m_H$ is related to some internal *yet unknown unit* for a bound 2-unit charge system, much like ±1 are the *known* units for algebraic numbers. *Conventionally, the handedness of algebraic numbers is considered as virtual or imaginary. To work with handed numbers, complex numbers must be used* [9]. *If so, there is an analogy between virtual handedness of algebraic numbers, expressible with ±i, and the real handedness for physical numbers, expressible with ±$m_e/m_H$* [9b]. Strictly spoken, one only needs *one axis* with *one origin* (or 1D) *to discuss algebraic numbers*. If we want to go over to *2 axes and origins* (a plane or 2D) or to *polar coordinates* for these same numbers, we introduce an extra (superfluous?), degree of freedom: a (*virtual*) handedness for numbers or, eventually, an imaginary chiral characteristic [9].

If the Bohr electron-proton model were valid, then Bohr unjustly suppressed an algebra related to left/right or mirror symmetry, exactly as we proved recently [6,7,9]. *Internal* algebra is suppressed with introducing charge invariance for the attractive Coulomb law. In terms of a lepton-baryon model in a positive mass world, Bohr overlooked *the internal asymmetry connected with the constitution (mass) of the baryon, needed to describe the H species by means of a proper unit*. In fact, *with infinitely heavy baryon mass*, the 2 reduced masses for H in the +world converge. But in this case, reduced mass makes no sense as it goes to zero. A trivial conclusion is that a matter-antimatter asymmetry must be most pronounced for H, where (relative) baryon mass is minimal.

For a 2-fermion Coulomb system, the mass-conjugation we wanted for the matter-antimatter (a)symmetry is finally confined to an internal algebra for the classical reduced mass. This *algebraic recoil theory* (ART) is the basis of a chiral field theory, where matter-antimatter asymmetry is a left-right or mirror asymmetry. As we will show below, reconsidering the Bohr model for H in terms of a Coulomb law reveals the same refinement with respect to charge conjugation.

*The simplicity of the effect of a zero mass perturbation does not prohibit the generation of some interesting consequences in the framework of classical physics, without having to recur to bound state QED: only number theory is involved.* We already remarked that all derivations above remain valid, whether or not we assign a charge to particles *subject* to a zero mass perturbation. In particular, all particles $m_x$, $m_e$, $M_p$ and $m_H$

can -in theory- be considered as *electrically neutral particles*, without affecting the above results. If true, we should rethink, as mentioned above, the Coulomb attraction (see below and [9]).

We verify that legitimate approximation +1001-1 for +1000, physically forbidden in the positive world, becomes legitimate in the positive world to describe number –1000 with –1001+1. *Only the units, ±1, are the decisive factor.*

The intimate connection of the above approximations for algebraic numbers with number π, and especially angle ½π for a mirror plane [6], is now visible. The result of applying a zero mass perturbation for a particle having non-zero mass is as generic as obtaining a value for π in the way Archimedes did, i.e. by means of *inscribing and circumscribing polygons with equivalent sides for a unit circle*. Inscribed (smaller) polygons lead to π from the lower side (1000-1=999), circumscribed (larger) polygons from the upper side (1000+1=1001). This analogy between algebraic number theory and approximations for number π is treated further below (Section 3.3), since there is an important *asymmetry* between in- and circumscribed polygons too, connected with a hypotenuse and with axes.

*2.3. Known problems with recoil and errors in QED*

The history of incorporating recoil in bound state QED is a long and delicate one [1,2,8] but need not be repeated here. We refer to 2 important remarks since we suspect chiral behavior in the electron-proton interaction may be linked with recoil, a classical *static* effect.

First, Erickson [8] found odd recoil effects in QED theory for the Lamb shift. Erickson wrote: '*This should constitute a strong warning that mass dependences can be rearranged in a number of useful forms and the success or utility of one form should not be taken as denying the possibility of another form also being successful or useful*' [8]. The matter is discussed in depth below and in [9], since Erickson actually meant here *that a major uncertainty in bound state QED is how to formulate exactly/analytically the mass dependences or, in fine, recoil $m_e/M_p$ if not modified recoil $m_e/m_H$*. Recoil, i.e. a classical static mass effect, is at the roots of our chiral theory above. We will now have to reinterpret Erickson's QED results for atom H [8] in terms of recoil (see Section 4) along the lines set out in previous reports [6,7,9].

A second remark is more general and points to a large error in bound state QED. Recoil enters the Sommerfeld-Dirac H energy expression as [1]

$E_{n,j}(H) = \mu c^2/\sqrt{1+(\alpha/(n-\varepsilon))^2]} - \mu c^2$

wherein $\epsilon = j+\frac{1}{2} - \sqrt{[(j+\frac{1}{2})^2-\alpha^2]} \approx \alpha^2/(2j+1)$, in which j is the total angular momentum. Here, Cagnac et al. [1] justly remark that the above Sommerfeld-Dirac expression *has no exact theoretical justification* (sic) *but it is rather simple*. In fact, recoil as used in this famous expression does not make sense [1]: there is no valid reason to subtract $\mu c^2$ instead of $m_e c^2$, since the unbound free electron (mass $m_e$) must no longer be connected with proton mass $M_p$.

If this were an error indeed, it is an extremely large one since $(m_e-\mu)c^2=(m_e/m_H)m_e c^2$ is about 5 times *larger* than Rydberg $\frac{1}{2}m_e\alpha^2 c^2$. How to solve this problem is shown elsewhere [20]. Moreover, the concept of reduced mass makes no sense either in a relativistic context [1].

*The conclusion of Section 2 is that we detected a matter-antimatter asymmetry for the 2-unit charge bound Coulomb system, the electron-proton system, which is not covered by Bohr theory.*

For the same system H, a unitary world-antiworld-description in terms of particle masses leads to an algebraic quadruplet, whereas, in terms of particle charges and Coulomb's law for the same system, it can only be a doublet. This difference provokes, in whatever world we choose, always a splitting in terms of the mass description, which is doomed to be invisible in a Bohr Coulomb description.

*This reduces to a difference in symmetry for the same system, pending the description, and differences in symmetry (presence or absence of a symmetry) are at the hart of differences between achiral and chiral behavior* [9].

We can now proceed analytically with the above asymmetries, absent in Bohr as well as in bound state QED theory. We must first find out whether or not these are hidden in Bohr theory, without recurring to bound state QED. Since chiral behavior is preponderant in naturally observed macrostructures, we must verify how chiral behavior works at the micro-level in nature, an atom.

**3. Consequences: how to get at a chiral field theory?**

Although the presence of charges on all above particles is not critical, it is obvious that for a bound 2-unit charge Coulomb system, the combined effect of world and anti-world is generating *a quadruplet of reduced masses*, instead of the single one in Bohr and QED theory. The maximum difference possible for a Coulomb 2-unit charge system is 2 (doublet ± or repulsion and attraction). The 4 reduced masses for the same 2-particle system in a mass-based world-antiworld description are $\mu_w$ (w=1 to 4) and are correlated 2 by 2 in different ways, pending the algebra chosen.

For 2-unit charge systems and in units of conventional reduced mass $\mu_{Bohr}$, the correlation in terms of their in-(I) and external (E) algebra is given in Scheme 1.

*Scheme 1. Correlations in the quadruplet of reduced mass for 2-unit charges (in units $|\mu_{Bohr}|$)*

| I\E | + | - |
|---|---|---|
| + | $-(1+2m_e/M_p)$ | $+(1+2m_e/M_p)$ |
| - | 1 | -1 |

A charge-based algebra for a 2-unit charge Coulomb system is *a doublet* ± like for algebraic numbers and their exact external parity. The mass-based algebra for a particle, perturbed with zero mass, is *generic quadruplet*

$$\mu' = \mu_w/m_e = \pm(1 \pm m_e/m_H) \qquad (16a)$$

As indicated in [6,9], (16a) shows that in a positive world baryon mass in the reduced mass recipe is either $m_H - m_e = M_p = 1836.1526675\, m_e$ as in Bohr theory or it is $m_H + m_e = 1838.1526675\, m_e$ in a chiral field theory. *The metaphor here is that the corresponding quadruplet for 2 conjugated complex numbers is $\pm(1\pm i)$, where we have left the axis (1D) for a number and replaced the axis with a plane (2D) for describing the same number* [9]. Representing $m_e/m_H$ by $1/x$, the generalized quadruplet is of form $\pm(1\pm 1/x)$, exactly of the type for the molecular Coulomb quadruplet in 4-fermion systems like molecules [5]. Let us look at further consequences of Scheme 1 and, in particular, to the *doublet-quadruplet* or *charge-mass* descriptions of the very same system H.

An *algebraic recoil theory* (ART) leads to the rationale behind mirror or matter-antimatter symmetry, which, for H, can be reproduced by means of two physically different *structures*, the *physical unit* being modified recoil $m_e/m_H$ [9].

*3.1 Is reduced mass algebraic (ART) and is it a continuous or a discrete variable?*

Having established the measure or unit for discrete mirror or left-right symmetry, the next problem is to verify if chiral behavior can vary and how this can be described analytically. The first most evident option is to consider p in (2b) as a parameter.

Even if p in (2b) were variable, the perturbation is still of zero mass type. We get a more generalized reduced mass quadruplet

$$\mu' = \mu_w/m_e = \pm p(1\pm pm_e/m_H) \qquad (16b)$$

instead of (16a), where only p=1 can produce a non-algebraic Bohr result. Generalized reduced mass (16b) in a positive world varies with p and has an extremum for a critical p-value equal to

$$p_{crit} = \tfrac{1}{2}m_H/m_e = 918.57633 \qquad (16c)$$

a rather strange result, we use elsewhere [20]. Inserting (16c) in (16b) gives extremes $\mu'_{crit}$

$$\mu'_{crit} = \pm\tfrac{1}{2}(m_H/m_e)(1\pm\tfrac{1}{2}) \qquad (16d)$$

$$\mu_{w(crit)} = \pm\tfrac{1}{2}m_H(1\pm\tfrac{1}{2}) \qquad (16e)$$

These correspond with

$$\mu_{w(crit)+} = \pm\tfrac{3}{4}m_H \qquad (16f)$$

$$\mu_{w(crit)-} = \pm\tfrac{1}{4}m_H \qquad (16g)$$

The ratio of the absolute values in (16f) and (16g) is 3. Instead of classical $\pm\tfrac{1}{2}m_H$ obtained *after achiral bisection of $m_H$*, different reduced masses are obtained, related to $m_H$ only, not to $m_e$ as in Bohr theory. We return to this generalization elsewhere [20]. In [9], *we said that, in Bohr theory or in QED, it is strange that mass $m_H$ does not show in the H-energy expressions, except for a small recoil correction.*

In the constraints of the present paper, result (16b) suggests continuously varying reduced mass, a consequence of the generalized zero mass perturbation for H and H̲.

If true, the possibility that *2-unit charge electron-proton system in a single world* may *fluctuate* between an *atomic-* and an *anti-atomic* state (and vice versa) cannot be excluded. This is the point we made in [6,7,9] in the context of chiral behavior in atom H. The different atomic- and anti-atomic-states of H can be thought off as back and front of the same neutral species H, which must always be differently handed. To make sense, *this means that species H must be divided internally to get at a mirror effect* (inversion in 1D). *The only generic characteristic of a mirror (plane) we know with absolute certitude is that, in 3D, its plane is confined to an angle of exactly 90° or ½π*. But an internal division of species H produces a fundamental problem with the Coulomb electron-proton attraction. Indeed, a Coulomb attraction -1/r only depends on the length of separation r between 2 charges: it is completely insensitive to *the internal division of the separation between the charges, its internal (a)symmetry, whereby a mirror plane at 90° may be involved* [9]. In other words, in 1D, the Coulomb law is insensitive to the position of the origin on the axis (the inversion center) or, it depends solely on the total length of a line segment, wherever situated on the axis.

*3.2 The 2-unit charge separation in the achiral Coulomb field and its chiral behavior*

Confirmation of a continuously varying reduced mass (16b) is easily obtained with Bohr theory, without QED. The equivalent of a formulation in terms of reduced masses is given by the separation $r_H$ for the 2-unit charges. We discuss the positive world solution first and use the Bohr reduced mass (4) to illustrate the procedure. We rewrite (4) as

$$\mu_1(+,-) = \mu_{Bohr} = 1/(1/m_e + 1/M_p) \qquad (16h)$$

Total energies are $m_e c^2 = a_e e^2/r_e$ and $M_p = a_p e^2/r_p$, $a_e$ and $a_p$ being scale factors. Using these in (16h) and putting $a = a_e = a_p$, we get the Bohr equilibrium condition for the 2-particle or pseudo one-particle rotator system

$$\mu_1(+,-) = \mu_{Bohr} = (1/a)(e^2/c^2)/(r_e + r_p) \qquad (16i)$$

This describes the classical transition from kinetic energy to Coulomb energy, as in the equilibrium condition used in Bohr's model. This proves how ART leads to an, otherwise seemingly irrelevant, internal algebra for the Coulomb field, our basis of a classical chiral field theory. Leaving out scale and conversion factors, we have in general *discrete* relationship: $\mu(\pm,\pm) \sim (\pm)1/[r_e(1 \pm r_p/r_e)]$.

Unlike (16b), discreteness disappears with classical physics. The classical *positive* solution for separation $r_H$ in the Bohr model is

$$r_H = r_e \sqrt{[1 + (r_p/r_e)^2 - 2(r_p/r_e)\cos\theta]} \qquad (17)$$

where $r_e$ and $r_p$ are the respective separations of the charges from the center of mass of the system and where $\theta$ is the angle between the two vectors $r_e$ and $r_p$ [7,9].

To go over to a notation for (17) in terms of particle masses, we use Bohr's classical reduced mass recipes

$$m_e r_e = M_p r_p \qquad (18a)$$

$$m_e/M_p = r_p/r_e \qquad (18b)$$

Introducing (18b) in (17) gives

$$r_H = r_e \sqrt{[1 + (m_e/M_p)^2 - 2(m_e/M_p)\cos\theta]} \qquad (19)$$

The suppression of the *internal algebra* in (19) by Bohr is easily typified classically: *Bohr did not allow for variations in θ, which he fixed at θ=π*. This is just the *anti-parallel configuration* of charges or of $r_e$ and $r_p$, which forbids the *parallel* one with $\theta = 0$ (or $2\pi$) along the line between the 2 unit charges

+e and –e (determining the *achiral* Coulomb field). With a number metaphor, Bohr only allowed to assess number +1000 with number +999+1, not with number +1001-1.

A variable θ produces different extremes (asymptotes) for linear alignments in *positive* $r_H(\theta)$:

a. $\theta_{Bohr}$: $r_H(\pi) = r_e + r_p = r_e(1+m_e/M_p)$ (20a)

b. $\theta=0$ $(2\pi)$: $r_H(0) = r_e - r_p = r_e(1-m_e/M_p)$ (20b)

These values are internally sign-correlated with those for reduced masses in Section 2.

In contrast with the discrete generic quadruplet of reduced masses (16a), (19) now gives, in first approximation, *a continuous quadruplet* [7,9]

$r_H/r_e \approx \pm[1 - (m_e/M_p)\cos\theta]$ (21a)

in function of angle θ. Circular function cosθ in (21a) has taken over the role of the internal algebra or the internal ± sign of the generalized reduced mass. Results (16a) and (21a) can be made consistent by assuming p in (2b) is not exactly equal to 1 and that it is variable too, see (16b), but this would require rescaling by p also (see Section 3.1 and [20]).

Applying (21a) to Bohr theory extends (*and refines* [6,9]) the Bohr energy formula to

$-E_{nH} \approx (R_\infty/n^2)[1+(m_e/M_p)\cos\theta]$ (21b)

showing classically that the extended or running Rydberg $R_H(n)$ in $1/n^2$ theory may fluctuate between two extreme values [6]. With (21b) $R_H(n)$-values are $R_H(n) \approx R_\infty[1+(m_e/M_p)\cos\theta]$ [7]. Further, expansion in the cosine also gives $R_H(n) \approx R_\infty[1+(m_e/M_p)(1-\tfrac{1}{2}\theta^2)]$, quadratic in angle θ [7,9]. *This angle must be assessed analytically, a difficult but not insoluble problem as shown in* [9b].

Alternative but classical descriptions (19) and (21b) literally show that a transition from one state to the other requires *crossing a perpendicular configuration with $\theta=\tfrac{1}{2}\pi$ or $90°$, the characteristic or generic angle for a mirror plane*. Conceptually, there is nothing wrong with an extended but still classical Bohr theory, i.e. letting angle θ vary, to understand running Rydbergs, as found recently [6,7,9]. *These unambiguously led to the concept of mirror symmetry for atoms* or to a CSB theory for atom H [6]. Intra-atomic symmetry is now *naively* understandable even with classical Bohr theory (21b) [6,7,9], without having to refer to bound state QED and its underlying Dirac and Einstein theories [9], where there are problems with recoil (see Section 2.3). A naïve *classical* interpretation of varying Rydbergs $R_H(n)$ [6], based upon (19) or (21) [7], does not require *new physics* [9].

The natural naïve explanation is indeed that $r_H$ (19), and hence $\cos\theta$, would vary with n also [7,9]. And if these fluctuations could be understood with classical physics [9b], the role or meaning of QED (and of STR) may even be questioned [6,7,9]. *We notice explicitly that original Bohr theory contains all ingredients to arrive at a chiral theory.* Only, these possible (*hidden*) extensions were suppressed unjustly although they are easily exposed [6,7,9]. *Obviously, there are obstacles to refine original Bohr theory in this naïve way otherwise results* [6,7] *would have been obtained earlier. What are the difficulties with this extension of Bohr theory* [9]?

3.3 *Difficulties with extended Bohr model (21b) for the 2-unit charge separation and chiral behavior*

(a) The correlation between internal algebraic recoil theory for μ and variable charge separations $r_H$ is not one by one, since (20b) suggests a composite particle (baryon) with mass $M_p - m_e$. Mass-conjugation in the context of chiral behavior must be interpreted with $m_H \pm m_e$ (16a) rather than with $M_p \pm m_e$ [9].

(b) With 2 reduced masses allowed in the same world, 2 Rydbergs are created for the same system H. This important difficulty is treated separately in Section 3.4.

(c) For an electron-proton attraction, an *internal division* of $r_H$ in $r_e$ and $r_p$ *does not affect the magnitude of the Coulomb force*, which is achiral or, at least too symmetrical [9] (see also below).

(d) This division generates *two monopoles* instead of a *dipole* for the bound 2-unit charge Coulomb system, *which must be avoided.* How to do this using a 4-fermion model for perturbed H, is explained in [9].

(e) An important difficulty is *the direction of the Coulomb field* in the hypothesis that angle θ is not a constant. With varying θ, the *direction* of the field between 2-unit charges will show small fluctuations in the absolute reference frame, which are difficult to understand. For a pair of conjugated charges, the direction of the field is from one particle (+) directly to the other (-) in a straight line (a dipole). Therefore, fluctuations in θ as in (19) mean that *both strength and direction of the 2-unit charge Coulomb attraction fluctuate*, which does not make sense in classical physics. *In a unitary rectangular triangle* ABC, with A charge +e, B origin O and C charge -e, hypotenuse AC ($r_H$) gives a maximum deviation of the field's direction from the axis. The strength of the perpendicular configuration depends on $1/\sqrt{[1+(m_e/M_p)^2]}$ since $\cos 90° = 0$. *The (minor) difficulty with the extension of*

*Bohr theory generates a (major) conflict with the framework of STR, the Special Theory of Relativity* [9b]. Although in the linear classical alignment as in Bohr theory there is no direct connection between relativity and recoil [1], the possibility that field directions are no longer on an axis generates extra problems.

(f) In fact, *if the direction of the field in the perpendicular state* were decisive to describe the energy of the electron-proton system, we get, using $u=m_e/M_p$, higher order terms of type

$$1/\sqrt{(1+u^2)}=1- \tfrac{1}{2}u^2 + \tfrac{3}{8}u^4 -\ldots$$

not only (see Section 4.3) of the same form as those in a Dirac-Einstein relativistic expansion in $\alpha^2$ [9]. *With classical physics, the use of a perpendicular model for scaling a uni-axial Coulomb field effect, a common practice in STR, cannot be understood* [9]. In addition, the *symmetrical* counterpart energies of type

$$1/\sqrt{(1-u^2)} =1+ \tfrac{1}{2}u^2 + \tfrac{3}{8}u^4 +\ldots$$

can never be obtained and even cannot make sense, since, with reference to STR equations, they are virtual or imaginary: only squared product $iu$ can give $-u^2$. *However, exactly this latter type of energy would confine the inter-particle separation to an axis, as it should, and not to the hypotenuse* [9]. We can now return to the analogy (Section 2) between the determination of $\pi$ and the zero mass perturbation for H. A circumscribed polygon gives larger rectangular triangles (formed by a bisected side and the center of the unit circle) than the inscribed polygon. For a *circumscribed polygon* of n equal sides with length $2L_c$, all 2n rectangular left and right handed triangles depend on hypotenuse $h=\sqrt{(1+L_c^2)}$. For an *inscribed polygon* with sides of length $2L_i$, triangles have a hypotenuse equal to 1. This means that one of their sides s (an axis) has length $s=\sqrt{(1-L_i^2)}$. *There is no reason to suppose that using a circumscribed polygon based upon $\sqrt{(1+L_c^2)}$ is the only real or allowed method to determine $\pi$ and that using an inscribed polygon based upon $\sqrt{(1-L_i^2)}$ is a virtual or forbidden method.* The only difference between the 2 Archimedean $\pi$-approximations is that one is bound to the circle from the *inside* (in 1D: N-1, *the inner world*), the other from the *outside* (in 1D: N+1, *the outer world*). With this model, the STR framework is related only to circumscribed polygons to determine $\pi$ and refutes the other procedure.

This seemingly trivial matter is of fundamental interest since it would reduce the variations in $\theta$ and the corresponding variations in strength and direction above to a variation in strength alone, *leaving the direction of the internal field unchanged and confined to an axis*. The full discussion is given in [9] but in the context of this work it is useful to state that the alternative to separation (19) derives from what we called an *engine model* [9]. This uses the 2 other angles η and χ in the same triangle ABC with θ=π-(η+χ) [9]. We obtain for an axis (a field)

$r_H = r_p\cos\eta + r_e\cos\chi$

instead of (19) [9]. Since[2] $r_p\sin\eta = r_e\sin\chi$, we get

$r_H = r_p\cos\eta + r_e\sqrt{[1-(m_e/M_p)^2(1-\cos^2\eta)]}$

which leads to $1/\sqrt{(1-u^2)}$ for the perpendicular case where η=½π or 90°. An *engine model based formula* for particle separation $r_H$ is different from Bohr's (19) [9b]. It secures that $1/\sqrt{(1-u^2)}$-type energies for perpendicular configurations are perfectly allowed, whereas, in the framework of STR, these can only be imaginary. A conceptual advantage of an engine model is that *a real perturbation can be understood in terms of starting an (hydrogen) engine* or *of fluctuations of an origin.*

Alternative engine formulae correspond with a fluctuating separation for the 2-unit charges too *but its direction remains confined to a single axis, as it should in a field theory.* Here, we see why and how the framework of Special Theory of Relativity is not an absolute one: perhaps, it is only of *relative* importance [9b]. With standard engineering mechanics interesting results are obtained for an engine in perturbed H, among which an alternative for cosθ above [9b].

A consistent generalization of Bohr H theory is important for understanding the laws of nature (see Introduction). In Section 3.3, we have shown that these theoretical consequences are of general interest for the 2-unit charge interaction and that their analytical formulation is unambiguous. *The conflict with STR is easily solved classically.* Now, we must test the above refinement of Bohr theory, possible with (19) and (21b) and with an engine model [9].

*3.4 Two different Rydbergs for atom H?*

Before doing so, we return to problem (b) in Section 3.3. *Allowing for parallel and anti-parallel structures for 2 charges in H is a substitute for matter-antimatter asymmetry.* With Bohr theory and p=1 in (2b), the problem is that *2 different discrete Rydbergs* are generated for H, given by

$R_H(+,-) = R_\infty (1-m_e/m_H) = R_\infty /(1+m_e/M_p) = R_H(Bohr)$ (22a)

$R_H(-,+) = R_\infty (1+m_e/m_H)$ (22b)

---

[2] Stevin, born in Bruges (Flanders) in 1548, used this formula to refute a *perpetuum mobile* in a way much admired by Feynman. Stevin concluded *Wonder en is gheen wonder*, which means: *A miracle is not a miracle*, *Magic without magic* (the title of Wheeler's album) or *Anomalous is not anomalous*, a QED-inspired translation [9].

corresponding with a large difference, not less than (2.59=) 118 cm$^{-1}$, if R$_\infty$≈109,737.3 cm$^{-1}$. *There is no experimental evidence for such large gap.* At first sight, a matter-antimatter asymmetry and 2 largely different Rydbergs (22) seem to be falsified by the H line spectrum, as remarked previously [9].

Despite this, quadruplet (21) suggests that, in a given world, there may nevertheless be a *continuous transition* between 2 Rydbergs, or 2 different reduced masses for neutral H. *If, at first sight, the existence of 2 largely different Rydbergs (22) contradicts observation, the challenge is to find out whether or not the line spectrum of atom H is consistent with a continuous transition between any 2 Rydbergs.*

We must decide on the more important aspect of CSB theory or ART for atom H: is it
(i) the existence of *2 different Rydbergs with a gap of 118 cm$^{-1}$* using the Bohr H model and R$_\infty$, an *external criterion*, or
(ii) the existence of *continuously varying Rydbergs* with an extremum at exactly ½π or 90° [6], *an internal criterion*, without fixing the difference between Rydbergs at 118 cm$^{-1}$ ?

This situation is comparable with asymptotes for circular functions pending the value of the angle. To go from asymptote +1 at angle 0° to asymptote –1 at 180° or π for a cosine, the zero asymptote at 90° or ½π must be crossed. For H, there may be something wrong with scaling asymptotes [9]. In fact, the 2 Rydbergs for H are conceptually related to asymptotes +1 and –1 for a cosine. *The internal mechanism, governed by a critical angle of 90° or ½π, can never be avoided, whatever the values assigned to external asymptotes. Internal constraint (ii) is more restrictive than external constraint (i).* The difference between the asymptotes (Rydbergs) is mainly a question of scaling [9]. Of course, there must be a relation between chiral and external asymptotes or Rydbergs, as we will show in [20].

*3.5 The detection of different Rydbergs and a CSB theory for atom H [6,7,9]*

Given their importance in the context of matter-antimatter (a)symmetry and constraint (ii) in Section 3.4, variations in R$_H$ in function of quantum number n can indeed be detected with running Rydbergs, defined as [6,7,9]

$$R_H(n) = -E_{nH} \cdot n^2 \qquad (23)$$

Running Rydbergs (23) cannot but describe the extra symmetry, *if any*, contained in a landscape for H *beyond* Bohr 1/n$^2$ theory, giving $-E_{nH}=R_H/n^2$. R$_H$(n) illustrates quantitatively the algebra (ART) overlooked in Bohr theory.

$R_H(n)$ (23) in [6], as well as a Mexican hat or double well curve for H in [7], show that a continuous transition between Rydbergs is very consistent indeed with the H line spectrum for ns½ singlet states. This proves *indirectly* that 2 different H Rydbergs exist (constraint (i) in Section 3.4): *only their difference is not equal to 118 cm$^{-1}$ but it is much smaller* [6,9]. The problem can be solved with rescaling the Rydberg [9,20]. Errors in Bohr theory are of order $10^{-7}$-$10^{-8}$, but varying Rydbergs give errors of only $10^{-11}$, a factor 1000 better (see below and [9]).

In addition, we showed that in H there is an internal mirror plane exactly at its generic angle $n_0=½\pi$ [6], where the maximum Rydberg is 109,679.352282 cm$^{-1}$ [6]. This is different from $R_H$ and from the NIST-value [6]. Only a *harmonic* Rydberg separates hydrogen- and antihydrogen-states exactly at $n_0=\pi$ in the Mexican hat curve [7]. The Bohr Rydberg $-E_{1H}$ is 109,678.773704 cm$^{-1}$ [6,8], the NIST value of 109,677.58553 cm$^{-1}$ applies to $-E_{\infty H}$ [6]. Maximum and minimum Rydbergs are separated by *1.7667 not 118 cm$^{-1}$* (see also [20]).

Earlier [6,7], we interpreted these results as a chiral symmetry breaking effect (mirror symmetry) without offering an explanation. Here, we see that it refers exactly to the symmetry we introduced above as a working hypothesis to discuss a zero mass perturbation for atom H. *By extension, the matter-antimatter (a)symmetry is now identified quantitatively*, although not yet completely understood. However, we see how useful simple and classical approaches can be for a fundamental problem: (a)symmetry of matter and antimatter and an inherent internal mass-conjugation in a single world. *We must now find out quantitatively if recoil or better the ratio $m_e/m_H$ is really at the hart of a chiral symmetry breaking process in species H* [9], since this problem was not discussed in [6,7].

*3.6 Putting the matter-antimatter asymmetry of CSB theory or ART quantitatively to the test*

Despite its shortcomings at this level, our analysis provides 3 quantitative criteria to put our mirror or matter-antimatter asymmetry hypothesis, based upon recoil, to the test.

(i) The only generic characteristic of a mirror plane is its angle: 90° or ½π.

(ii) Using (19), a separation of contributions from world and antiworld to a scaled H line spectrum is governed by 2 different coefficients

$$\pm m_e/m_H = \pm 0.000544\ldots \qquad (24)$$

with $m_e/m_H=1/1837.1526675=0.0005443206$. The difference between 2 coefficients $+m_e/m_H$ and $-m_e/m_H$ is $2m_e/m_H$, discussed around (15), see also (21a).

(iii) As a consequence of the Mexican hat curve [7], symmetry in the complete n-set (n=1 to 20) for Lyman $ns½$ singlets [8] is broken exactly at $n_0=\pi$. This gives set A with 3 data points (n=1 to n=3) and set B with 17 (3<n≤20). Then, we must find out which set, A or B, belongs to the anti-atom domain in the positive world. The guiding line is negative coefficient $-m_e/m_H$ in (24), which is related to the conventional Bohr atom world, since, with $m_H=M_p+m_e$, we have $1- m_e/m_H = M_p/(M_p+m_e)= 1/(1+m_e/M_p)$ at the roots of the standard H Bohr model.

A *practical* constraint for A with 3 data points is that fitting is limited to order 2 (quadratic fit). This is the phenomenological procedure we adopt now but we must arrive at the analytical form of a suited linear 1D representation for H, a topic discussed in full in [9,20].

*3.7 CSB theory or ART: scaling the primary 1D field axis for matter- and antimatter-worlds*

With 19th century chiral behavior, 3D structures are stable, which implies a minimum of 4 particles [9][3]. To generalize left-right (a)symmetry, the 3D structure must be *folded* [9,21-23] to arrive, over an *intermediary* 2D-structure, to a linear 1D equivalent to quantify the uni-axial (static) field-effect, responsible for a left-right difference on a single axis (the uni-axial representation of a field) [9]. The equivalent is *unfolding an origin on an axis* and gradually unfold further to reach 3D [9]. The consequence for the inverse *unfolding* procedure is trivial: to arrive from a 1D to 3D structure, we must gradually go from 2 (1D), over 3 (2D), to at least 4 point-like particles (3D). *Therefore, if natural perturbed atom H shows left-right asymmetry, at least 4 fermions are required, which makes Bohr's model 2 fermions or 1 boson short*. We can solve this problem and avoid monopoles, see [9].

The intermediate planar 2D complex for which 3 particles are sufficient can be visualized by a bisected equilateral triangle, where one half is left-, the other right-handed. Without leaving the plane, the two halves can never be made to coincide, even when their surfaces are equal (*Lord Kelvin's definition of chiral behavior*). Alternatively, back and front of a triangle are always differently handed too: if a 4th particle is involved, the handedness of the 4-particle structure depends on the position of the 4th particle: in front or at the back of a triangle [9]. In any case, the triangle (3-

---

[3] In [9], we explain why CSB H theory requires 4 fermions instead of 2 in Bohr theory [6].

particle system or N=3) is critical to distinguish between symmetry and energy based models to describe the final 4-particle structure (see below and [9]).

In principle, a 2-unit (point-like) charge system is uni-dimensional or 1D and we suspect that, if we want to quantify all of the above, especially the coefficients (24), we are confined to use a linear 1D model based upon line segment $r_H$, best suited for this purpose. In fact, we need the best description for *2 point-like particles situated on a single axis and an achiral or chiral division of the line segment between the two* [9]. The problem was discussed in Section 3.3, point (e), with reference to the engine model, STR and Archimedean π approximations [9]. The important difference between uniaxial and biaxial models for field effects, thought to be uniaxial, is discussed more fully elsewhere, also in connection with STR, the Special Theory of Relativity [9] (see also Section 3.3).

*3.8 Difference between symmetry- and energy-models. Continuous chirality measures (CCMs)* [9,21-23].

There is a difference between a description of N-particle (-fermion) systems using a symmetry- and an energy-based model with Coulomb's power law [9]. *For symmetry-based models*, the positions of the N vertices of a structure with respect to the *center of the structure* are required. *For energy-based models* with Coulomb's 1/r law, *all separations* r between N particles are required. It is easily shown that the 2 models converge if and only if N=3 [9].

One consequence is that Coulomb's law for a bound 2-unit charge attraction (N=2) is always *achiral*[4] or (too) *symmetrical* [9]. If the line segment between 2-unit charges as in H is $r_H$, bisection using 2 equal charges for scaling gives two sign-conjugated line segments $+\frac{1}{2}r_H$ and $-\frac{1}{2}r_H$ of equal length (monopoles), meeting at the origin. Inversely, *the origin is unfolded in an achiral way if 2conjugated unit charges attract: there is left-right symmetry, determined by external algebra ± 1only (or ±½, if a different scale factor is used)* [9]. This *external* algebra secures the 2 halves of line segment $r_H$ can never be made to coincide *without leaving the axis* (Lord Kelvin's definition of chiral behavior in 1D). We said that for a bound 2-unit charge Coulomb system, *the description is always too achiral* [9]. In fact, for a result based upon the Coulomb law or $-e^2/r_H$, the internal symmetry of line segment $r_H$ *does not matter: it is overlooked/disregarded, as remarked above.* Up to this level, the analogy with fermion spin is obvious, especially since spin is defined with units (numbers) ±½. The above model with ±½$r_H$ is conceptually classical and more illustrative.

---

[4] Chirality means an absence of *a* symmetry. The closest achiral (more symmetrical) structure contains this symmetry element [9], an important aspect of chiral behavior [23].

In fact, when this very same line segment $r_H$ is divided *asymmetrically* or less symmetrically (*proportionally*), as it is in (19) for H due to classical static mass-bound field effects (18), less achiral, which literally means chiral, behavior sets in [9]. *Now, there is left-right asymmetry, the asymmetric part of the left-right symmetry being determined by internal algebra (16a), as explained above* (Scheme 1). *Generically, we say simply that the charge symmetry in the electron-proton system in H is broken by (a large) mass (difference).*

A full account is elsewhere [9] but the analysis here suffices to understand the procedure adopted in a CSB H theory. Avnir's group [21-23] showed that *the chiral behavior of observed stable chiral 3D structures can vary in a continuous way*. For chiral behavior, they use CCMs, *continuous chirality measures* [23], which allow dealing quantitatively with observed left-right transitions. So, *we already classified continuous quadruplet (21) as a quantitative CCM for atom H* [9b]. The prototype is a classical 19[th] century Walden inversion: *a transition from a left- to a right-handed stable 3D structure, consisting of at least 4 particles* [9b]. The reaction describing the Walden left-right inversion needs a *quadratic* description for reactivity parameters with an extremum at 90° (½π) for the mirror plane [21-23], *which is exactly the extremum point for running Rydbergs (23) for atom H* [6]. This similarity between experiments on classical (variable) chiral behavior of stable structures, known from the 19[th] century, and the running Rydbergs for H [6,7], led us to conclude that, eventually also in atom H, (variable) chiral behavior is detected [6,7,9]. We said [6] that the theoretical framework of bound state QED can not cope with classical 19[th] century chiral behavior: our formula above, are absent in bound state QED.

To quantify (21), we expect that only data corresponding with a linear 1D structure (on a single axis) will be able to confirm the predicted value of left-right coefficients ±$m_e$/$m_H$ (24) for H in the matter-antimatter (a)symmetry model introduced here. Finding the best 1D model is a problem. Taking the square root of Bohr's result or $\sqrt{(R_H/n^2)}$ gives linearity in function of 1/n but makes no reference to the mirror symmetry we detected [6,7].

We now proceed with Section 3.6, where the 20 data of the Lyman ns-singlet series [8] were divided in subsets A and B due to the critical n-value π in the Mexican hat curve [7]. We remark that, using running Rydbergs [6] with critical n-value ½π=1.57..., only Bohr's ground state for H (n=1) separates from all other states with n>1 due to an asymmetry, *a separation invisible in Bohr theory and in bound state QED as well.*

To solve the recoil related difficulties, exposed by Erickson [8] (see Section 2.3), we must make advantage of the only internal anchor available (½π) to rewrite one-electron energies for atom H in a linear form: *the mirror symmetry we detected* [6,7,9] *must be the basis of a linear H model.*

*3.9 Proving the matter-antimatter (a)symmetry: a phenomenological linear 1D approach for atom H*

To settle these problems with recoil [8], we *anchor* the H-data by means of variable π, the crucial issue in chiral field theory. For a linear representation of H symmetry breaking details, we restart the procedure with running Rydbergs $R_H(n)$ [6], governed by generic mirror plane parameter ½π [6]. A quadratic fit for each set A and B gives slightly different critical $n_{0A}$ and $n_{0B}$ values and harmonic Rydbergs $R_{harm}(A)$ and $R_{harm}(B)$.

*Chiral contributions* $CC_{nH}$ are positive [6,7]

$$CC_{nH} = R_{harm} - R_H(n) = a_s(1-n_0/n)^2 \qquad (25)$$

where $n_0$ is very close to ½π [6] and $a_s$ is an internal chiral asymptote [6,9]. For A and B, a linear 1D representation, *anchored by mirror plane reference value ½π* [6], is now easily obtained with the square root of (25) or

$$\sqrt{CC_{nH}} = (1-n_0/n)\sqrt{a_s} \qquad (26)$$

To get a measure as close as possible to the real situation, we compute an average $n_0$ with

$$n_0 = \tfrac{1}{2}(n_{0A}+n_{0B}) \qquad (27)$$

and use this to plot $\sqrt{CC_{nH}}$ from (26) versus

$$n_0/n-1 \qquad (28)$$

Next, we fit experimental $\sqrt{CC_{nH}}$ (26) versus (28) for A and B. This gives 2 data sets in the whole n-range (1 to 20) and will show how these depend on mean value variable (28). In this way, the linear 1D model (26) for H is anchored with ½π as we attempted. Then, we can determine the *linear characteristics* of the H Lyman ns-singlets for hydrogen- and antihydrogen-states and compare field properties of matter- with those of antimatter-states, expected to obey different recoil coefficients (24). Differences $\Delta_{nH}$ between data sets $\sqrt{CC_{nH}}(A_{theo})$ and $\sqrt{CC_{nH}}(B_{theo})$, plotted versus (28), lead to linear 1D coefficients, *also anchored with mirror angle ½π*. Results are in Section 4.

*3.10 The natural extension: a real non-zero or asymmetrical mass perturbation*

Using a zero-mass (mass-symmetrical) perturbation it is not difficult to consider the effects of a non-zero-mass or asymmetrical mass perturbation for H. In terms of H energy, the symmetrical perturbation (1) gives energy

$$E_H = +m_H c^2 \pm 0 \qquad (29)$$

A real perturbation (an asymmetrical mass perturbation) corresponds with different p-values for world and antiworld in the perturbing pair. The symmetrical case with the same p-value in either world or $\pm p(m_e - m_e)$ was discussed above. A variable p (Section 3.2) does not affect the internal symmetry of a zero mass perturbation.

Asymmetrical or non-zero mass perturbations (perturbations by kT, hν, atom…) imply different $pm_{xn}$-values for world and antiworld or at least $p_1 m_{x1} - p_2 m_{x2} \neq 0$. If $p_2=1$, as in Bohr theory, and $p_1>1$, the effect of a real perturbation in the real world is a rescaling (lowering by $1/p_1$) of the energy of the system to *a less stable form*. Rescaling for H in general is discussed in [20].

## 4. Results and discussion

*4.1 The appearance of world and antiworld chiral field coefficients $\pm m_e/m_H$ (24) in the H line spectrum*

As accurate data for a *complete set* from n 1 to 20 are not available (see Section 5), data are calculated as in [6,7] using Erickson's QED results [8]. Data are collected in Table 1.

For $R_H(n)$, the quadratic fit for the complete set of 20 data points was given in [6]. We repeat the result: $R_H(n) = -4.3674718717/n^2 + 5.5556169344/n + 109677.585534798$ cm$^{-1}$, giving $n_0 = 1.5722725$ and $R_{harm} = 109,679.352282$ cm$^{-1}$ [6].

For sets A and B we obtain slightly different values instead or:

(A)  $R_H(n)(A) = -4.3657644130/n^2 + 5.5533730093/n + 109677.5860953890$ cm$^{-1}$, giving $R_{harm}(A) = 109,679.352106$ cm$^{-1}$ and $n_{0A} = 1.5722929$ and

(B)  $R_H(n)(B) = -4.3726920038/n^2 + 5.5576673139/n + 109677.5854005220$ cm$^{-1}$, leading to $R_{harm}(B) = 109,679.3513416$ cm$^{-1}$ and $n_{0B} = 1.5735710$.

Average $n_0$, defined by (27), is $n_0 = 1.5729319$. Fits suffice to construct a ½π-related 1D presentation for atom H, conform our requirements above (see Section 3.9).

Fig. 1 gives the plot of $\sqrt{CC_{nH}}(A,B)$ versus $(n_0/n-1)$. The linear fits are

$$\sqrt{CC_{nH}}(A) = -1.3283735313(n_0/n-1) + 0.0005399081 \qquad (30a)$$

$$\sqrt{CC_{nH}}(B) = -1.3294271704(n_0/n-1) - 0.0005399910 \qquad (30b)$$

*Result (30) shows that the sets separate clearly as expected theoretically with (24)*. With NIST, we have $\pm m_e/m_H = \pm 0.0005443206$. For A, $n=n_0$ in (30a) gives $+0.0005399081$, within 1.02 % of the NIST-value. For B $n=n_0$ in (30b) gives 0.80 % deviation for $-0.000539991$.

*The average error (not shown here, but see [9]) for the 20 data with fits (30) is 0.77 kHz [9b], corresponding with only $2.7 \cdot 10^{-13}$ for 19 terms, seemingly sufficient to compete eventually with the accuracy, claimed by QED* [9]. We verify that a phenomenological CSB theory for atom H reproduces QED-results without loss of accuracy but without needing the QED hocus-pocus [9].

*To get finally at the matter-antimatter world anomaly, invisible in QED*, we simply subtract (30b) from (30a) to yield the differences $\Delta_{nH}$. The fit is

$$\Delta_{nH} = \sqrt{CC_{nH}}(A) - \sqrt{CC_{nH}}(B) = 0.0010536390(n_0/n-1) + 0.0010798991 \qquad (31a)$$

$$\approx 0.001063 n_0/n \approx 2(m_e/m_H)n_0/n \qquad (31b)$$

Fig. 1 includes the plot of $\Delta_{nH}$ versus variable (28). It is easily verified from results (30)-(31) that, differentiating between matter-world and antimatter-world using the H line spectrum reasonably reproduces theoretical coefficients $+m_e/m_H$ and $-m_e/m_H$ for each world, see (24), since $+m_e/m_H - (-m_e/m_H) = 2m_e/m_H$. We conclude that the theoretical world-antiworld asymmetry we anticipated is reasonably obeyed *experimentally*. *Analytical* result (31) is not covered *theoretically* by highly complex bound state QED [2] at all, where there are still difficulties with the role of mass or recoil, as remarked by Erickson [8] (see also Section 2.3).

Furthermore, the coefficient for set A is $+m_e/m_H$ and for set B it is $-m_e/m_H$. *This proves that set A with $n<\pi$ belongs to the antihydrogen or antimatter-world where baryon mass is $m_H+m_e$; B, with $n>\pi$, belongs to the conventional Bohr matter-world, where baryon mass is $m_H-m_e = M_p$.* Although this conclusive and detailed evidence could not be reached with the Mexican hat curve, detected earlier [7], this result is completely out of reach with present day QED. If all this were true, *the important term 1S-2S for natural perturbed H (critical for CPT and for the artificial antihydrogen projects at CERN /Fermilab) belongs to the antihydrogen-, not to the hydrogen-, domain* [7,9].

Reminding that $n_0$ in (27) and (31b) is close to $\frac{1}{2}\pi$ [6,7], we can rewrite (31b) as

$\Delta_{nH} \approx 0.001063 n_0/n \approx (m_e/m_H)\pi/n$ (31c)

a strange analytical result, *wherein the fine structure constant is absent*. In view of Erickson's remark on mass dependences in bound state QED [8], especially those related to the Lamb shift [6] (see Section 2.3), result (31c) is remarkable. Pending the correct value for the external asymptote, n-independent matter-antimatter asymmetry $n\Delta_{nH}$ seems to be solely a *matter* of recoil (or better $m_e/m_H$) and of number $\pi$. In an anti-matter world it is $+\frac{1}{2}\pi m_e/m_H$, it is $-\frac{1}{2}\pi m_e/m_H$ in a matter world. Matter and antimatter worlds are differently *scaled*, the ultimate basis of left-right asymmetry and of H chiral behavior in the real world [9]. All these pertinent consequences of the matter-antimatter asymmetry we introduced are invisible in the standard framework of QED. Moreover, the inconsistency with the use of reduced mass in QED (see Section 2.3) can be removed. We must conclude that QED cannot yet be validated as it stands, since, not only with its ambiguous use of recoil, it is unclear whereto QED is heading in the end.

Unprecedented results (30)-(31) are important too for electron mass $m_e$. With $m_e$ based upon Bohr's electron-proton model, NIST ratio $M_p/m_e=1,836.1526675$ applies. An anomaly may, however, occur on account of different scaling in world and anti-world. And, *as soon as it occurs*, it must have a value of $(m_H+m_e)/(m_H-m_e)-1=1,838.1526675/1,836.1526675-1\approx 2m_e/m_H \approx 0.0010892$ [9]. This result is discussed in Section 4.3.

*4.2 Energy differences between 2-unit Coulomb systems in world and antiworld*

With (31), we now *estimate* the energy differences between a 2-unit charge system in a world and in an antiworld to get an idea about the asymmetry. We use the running Rydbergs $R_H(n)_A$ in the anti-matter-world and those in the matter-world $R_H(n)_B$. Differences $\delta E_{nH}$, reflecting the matter-antimatter asymmetry in terms of energy, are given by

$\delta E_{nH} = [R_H(n)_A - R_H(n)_B]/n^2$ (32)

These *tentative* data are in the last column of Table 1. Fig. 2 shows the results in MHz.

This approximation is only tentative, as it is constrained by a 2nd order fit, by the small number of data in set A and by our limited knowledge of the subject at large. In fact, it may well be that our results are *in part* accidental due to the rather arbitrary definition of $n_0$ in (27). How to circumvent these limitations and uncertainties is shown elsewhere [20]. Nevertheless, *and within the* *constraints of our present approach*, world-antiworld asymmetry in H seems large for n=1 and decreases with increasing n. The meaning of the 4th order fit for (32) in Fig. 2 is not clear for the reasons mentioned above. The same applies for the values of coefficients in $1/n^2$, $1/n^3$ and $1/n^4$. Further work is needed [20] but we have a view on matter-antimatter (a)symmetry from Erickson's data [8].

*4.3 Unexpected external confirmation from the observed free electron mass*

An *anomaly* between matter- and antimatter worlds (see Section 4.1) must, *if it occurred*, be of order 0.0011, conform to (13) and (15) and their analytical equivalents $2m_e/M_p$ and $2m_e/m_H$.

Free electron mass $m_e$ can be assessed from its magnetic moment, which depends directly on $m_e$, the only variable in the expression for this moment. A long time ago, Dehmelt succeeded in measuring this moment, see [24], and found an *anomaly* of 0.0011592 (NIST). Schwinger [25] explained this anomaly with a first order correction term $½α/π=α/2π = 0.00116…$. Surprisingly, these 2 values are, in turn, close to the mass anomaly $2m_e/M_p$ (13). Inverse values $2π/α$ and $½M_p/m_e$ are 861 and 918 respectively, a small difference of 57 (6 %), which is too large in present day precision physics [9]. We already said that number $2m_e/m_H$, if really of basic importance for matter-antimatter asymmetries in nature, must be a fundamental constant of nature. Whether or not it is competitive with concurrent $α/2π$, the most important ingredient of bound state QED [2,9], remains to be determined.

The intimate affinity of $½α/π$ and $2m_e/m_H$ weighs heavily on our discussion of expansions in $(m_e/m_H)^2$ and STR expansions in $α^2$, discussed in Section 3.3 (f). We now see that not only the type of the expansions is similar but also the physics behind the expansions is very similar.

Conversely, the ratio of recoil and fine structure constant $(½m_H/m_e)/(1/α)=αm_H/(2m_e) ≈2π$ may be looked at as *a physical determination of π* [9], an issue we will raise again [20]. In this form, the analogy with *an Archimedean or mathematical determination of π* (see Sections 2 and 3) becomes apparent. If our analysis were valid, *the observed free electron mass complies with an antimatter-state configuration for a bound 2-unit Coulomb system in the matter world.*

Finally, the reference mass for a free electron is not a proton with mass $m_H-m_e=M_p$ but a baryon[5] with mass $m_H+m_e$ [9b]. Today, only $M_p$ is the basis of modern metrology [1]. In the present theory, exactly this same anomaly determines the world-antiworld difference for the so-called *simple* electron-proton system in atom H. QED uses corrections for this anomaly [1,2,8]: in CSB, the anomaly is not a correction as it is at the hart of chiral behavior in atom H.

---

[5] We will not speculate here on its relation with the neutron, which has mass very close to $m_H+m_e$.

*4.4 External confirmation from chemical bonds: atom-antiatom chemistry*

At the *intra-atomic level*, it is difficult to expose the left-right difference between the atom- and the antiatom-states since it is small (charge invariance of the 2-unit charge or electron-proton Coulomb attraction). At the *inter-atomic level*, its effect must be more pronounced. In fact, atom-atom or antiatom-antiatom Hamiltonians differ from the atom-antiatom Hamiltonian, since a parity operator (a switch) is needed to account for a charge inversion in the antiatom [5]. *A charge-related parity operator within a molecular Hamiltonian is absent in Heitler-London theory* [3] (see [5]). As stated in the Introduction, the stability of an atom-antiatom complex remains ambiguous despite many attempts to assess it [10-16]. We remind that 4-particle systems are insoluble anyway [5,9]. As shown before [5], *the main characteristic of atom-antiatom Hamiltonians is that nucleon-nucleon interaction is attractive instead of being repulsive* (as it is in Heitler-London theory [3]). This picture is not only very consistent with observed molecular PECs (potential energy curves) [5] but also with the general and intuitive idea, persisting since many a decade, that a molecular function should vary as $-1/R$, where R is the inter-nuclear separation, the bond length [5,9]. This shows that our results above, extracted from atomic line spectra, are confirmed by molecular band spectra [5,9].

It is remarkable that the chemical bond theory, closest to the concept of chemical atom-antiatom bonding, is old-fashioned naive *ionic bonding* of type $-1/R$ too [5,9]: *a nucleon-nucleon attraction* introduced in the 18th century (Davy, Berzelius...). *This concept was abolished almost immediately, but unjustly* [5,9], *after Heitler-London theory* [3] *was available.*

*4.5 Chiral field theory and metrology.*

An important consequence of ART or CSB theory for atom H is that the relation between the Rydberg for atom H, $R_H$, and its counterpart $R_\infty = R_H(1+m_e/M_p)$ has lost its uniqueness and so, its reliability for metrology [1]. This problem is so important for metrology (and the whole of physics) that it is far beyond the scope of the present report. We only want to point out that a metrology, even when based upon the anomalous mass of the free electron (see Section 4.3), is still susceptible of a relatively simple metrological treatment.

## 5. Conclusion

Classical physics allows understanding an internal mass-conjugation and its implicit matter-antimatter asymmetry in a single world, where only one kind of mass (+ or -) is allowed. A mass-based physical description for a 2-particle system provides with more algebra (symmetry) than a 2-particle Coulomb law, *if the internal symmetry of the separation between the 2 charges is overlooked. To arrive at additional symmetry, a Coulomb field must be assigned a gauge or the (field) axis must have an origin.* This result must lead to a more general approach [20].

Validating bound state QED today is indeed premature as argued in [6,7,9]. If level energies are anchored with mirror characteristic ½π, we get a quantitative exposure of the role of recoil for 2-unit charge systems, impossible or at least uncertain in bound state QED [8]. The picture for atom H arrived at with CSB is completely different from that given by QED.

To go further with this work *the complete Lyman $ns^{1/2}$ singlet series from n=1 to about 20 must be measured as precisely and as soon as possible.* The precision for term H1S-2S [26] is a guideline for this urgently needed fieldwork. The *experimental* set of 20 H terms available [27] has uncertainties of 3 MHz (or 0.0001 cm$^{-1}$), too large for matter-antimatter asymmetries detected now.

A major goal of CERN/Fermilab H̲ experiments is measuring the H̲ 1S-2S-transition as a test for CPT and QED. We repeat [6,7] that this term was already measured accurately [26]. Seeing the present results and [6,7], *this measured term H1S-2S [26] belongs to the H̲-domain in the real world.*

When looked at in this perspective, *the mass of the free electron is not that anomalous. With Stevin, we argue that, in this case, anomalous is not anomalous*, which typifies the general theme of our reports on the H line spectrum and on QED in particular [9,20].

Finally, the historical Lamb and Retherford 1950 remark that *the Coulomb attraction between electron and proton is not well understood* [28] was visionary. *The 2-unit charge Coulomb attraction was considered as too achiral, as 2 equal charges reside on particles with different mass. Static mass effects like in a balance* [9] *secure the H electron-proton attraction is chiral rather than achiral.*

*Acknowledgement*

I am in debt to G. Erickson and L. Wolniewicz for correspondence and to C. Chantler for useful remarks at PSAS2002 (St. Petersburg). Discussions with M. Tomaselli at the Wigner Centennial Conference (Pecz, 2002) were very inspiring.
*References*

[ 1] B. Cagnac, M. D. Plimmer, L. Julien and F. Biraben, Rep. Prog. Phys., **57**(9), 853 (1994)

[ 2] M. I. Eides, H. Grotch and V.A. Shelyuto, Phys. Rep., **342**, 63 (2001); hep-ph/0002158.

[ 3] W. Heitler and F. London, Z. Phys., **44**, 455 (1927)

[ 4] G. Van Hooydonk, phys/0007047

[ 5] G. Van Hooydonk, Spectrochim. Acta A, **56**, 2273 (2000); phys/0003005, submitted

[ 6] G. Van Hooydonk, Phys. Rev. A, to be published, CPS: physchem/0204004

[ 7] G. Van Hooydonk, CPS: physchem/0205004, submitted

[ 8] G.W. Erickson, J. Phys. Chem. Ref. Data, **6**, 831 (1977)

[9a] G. Van Hooydonk, *Anomalous is not anomalous (Stevin). Part I*, CPS: physchem/0204008

[9b] G. Van Hooydonk, *Anomalous is not anomalous (Stevin). Part II*, CPS: physchem/0204009 (also available as supplements to CERN-EXT-2002-041 dd 20/04/02)

[10] D.L. Morgan and V.W. Hughes, Phys.Rev. A, **7**, 1811 (1973)

[11] W. Kolos, D.L. Morgan, D.M. Schrader and L. Wolniewicz, Phys. Rev. A, **11**, 1792 (1975)

[12a] E.A.G. Armour and V. Zeman, Int. J. Quant. Chem., **74**, 645 (1999)

[12b] E.A.G. Armour and C.W. Chamberlain, Few-Body Systems, **31**, 101 (2002)

[13a] J. M. Richard, Phys. Rev. A, **49**, 3573 (1994)

[13b] K. Varga, S. Fleck and J. M. Richard, Europhys. Lett., **37**, 183 (1997)

[14a] S. Jonsell, A. Saenz, P. Froelich, B. Zygelman and A. Dalgarno, Phys Rev. A, 6405 (5), 2712 (2001)

[14b] P. Froelich, S. Jonsell, A. Saenz, B. Zygelman and A. Dalgarno, Phys. Rev. Lett., **84**, 4577 (2000)

[15] M. A. Abdel-Raouf and J. Ladik, J. Phys. B, **25**, L199 (1992)

[16] D. Bressanini, M. Mella and G. Morosi, Phys. Rev. A, **55**, 200 (1997)

[17] W. Kolos and L. Wolniewicz, J. Chem.Phys., **49**, 404 (1968)

[18] L. Wolniewicz, J. Chem. Phys., **103**, 1792 (1995)

[19] J. Eades, F.J. Hartmann, Rev. Mod. Phys., **71**, 373 (1999)

[20] G. Van Hooydonk, *Anomalous is not anomalous (Stevin). Part III*, in preparation

[21] H. Zabrodsky, S. Peleg and D. Avnir, J. Am. Chem. Soc., **114**, 7843 (1992)

[22] H. Zabrodsky, S. Peleg and D. Avnir, J. Am. Chem. Soc., **115**, 8278 (1993)

[23] H. Zabrodsky and D. Avnir, J. Am. Chem. Soc., **117**, 462 (1995)

[24] H.G. Dehmelt, Science, **124**, 1039 (1956)

[25] J. Schwinger, Phys. Rev., **73**, 416 (1948)

[26] M. Niering, R. Holzwarth, J. Reichert, P. Pokasov, T. Udem, M. Weitz, T.W. Hänsch, P. Lemonde, G. Santarelli, M. Abgrall, P. Laurent, C. Salomon, A. Clairon, Phys. Rev. Lett. **84**, 5496 (2000)

[27] R.L. Kelly, J. Phys. Chem. Ref. Data, **16**, Suppl. 1 (1987)

[28] W.C. Lamb Jr. and Retherford, Phys. Rev., **79**, 549 (1950)

Fig. 1 Matter-antimatter field asymmetry: plot of $\sqrt{CC_{nH}}$ ($\square$ set A, $\circ$ set B) and $\Delta_{nH}$ ($\triangle$) versus $n_0/n-1$

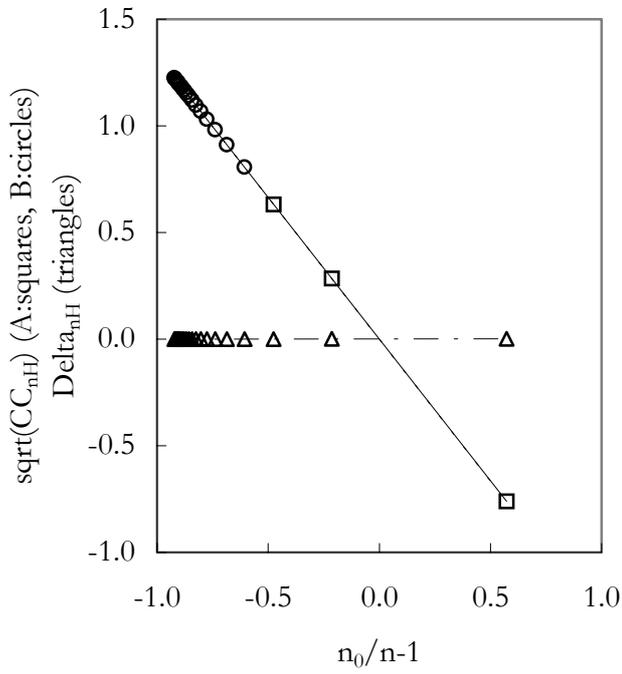

Fig. 2 Matter-antimatter asymmetry: plot of $\delta E_{nH}$ (MHz) versus $1/n$

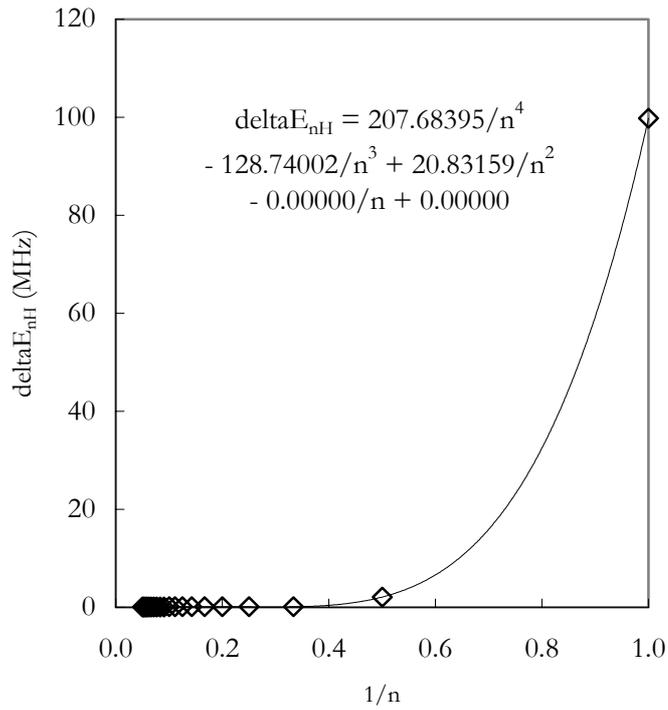

*Table 1*. Data to expose matter-antimatter asymmetry in the electron-proton attraction.

| n | $-E_{ns1/2}$ (cm$^{-1}$) [8] | A: $R_H(n)$ (cm$^{-1}$) | B: $R_H(n)$ (cm$^{-1}$) | $n_0/n-1$ | A: $\sqrt{CC_{nH}}$ | B: $\sqrt{CC_{nH}}$ | $\Delta\sqrt{CC_{nH}}$ | $\delta E_{nH}$ (MHz) |
|---|---|---|---|---|---|---|---|---|
| 1 | 109678.77370400 | 109678.77370400 | | 0.572932 | -0.760528 | | 0.001684 | 99.775525 |
| 2 | 27419.81783520 | 109679.27134080 | | -0.213534 | 0.284193 | | 0.000855 | 2.095642 |
| 3 | 12186.55023720 | 109678.95213480 | | -0.475689 | 0.632433 | | 0.000579 | 0.110472 |
| 4 | 6854.91884539 | | 109678.70152624 | -0.606767 | | 0.806111 | 0.000441 | 0.101677 |
| 5 | 4387.14088090 | | 109678.52202250 | -0.685414 | | 0.910670 | 0.000358 | 0.135638 |
| 6 | 3046.62195040 | | 109678.39021440 | -0.737845 | | 0.980371 | 0.000302 | 0.142887 |
| 7 | 2238.33245130 | | 109678.29011370 | -0.775295 | | 1.030159 | 0.000263 | 0.136298 |
| 8 | 1713.72205915 | | 109678.21178560 | -0.803384 | | 1.067500 | 0.000233 | 0.124752 |
| 9 | 1354.05122143 | | 109678.14893583 | -0.825230 | | 1.096543 | 0.000210 | 0.112236 |
| 10 | 1096.78097442 | | 109678.09744200 | -0.842707 | | 1.119777 | 0.000192 | 0.100344 |
| 11 | 906.43020253 | | 109678.05450613 | -0.857006 | | 1.138787 | 0.000177 | 0.089623 |
| 12 | 761.65290399 | | 109678.01817456 | -0.868922 | | 1.154629 | 0.000164 | 0.080177 |
| 13 | 648.98217184 | | 109677.98704096 | -0.879005 | | 1.168033 | 0.000154 | 0.071937 |
| 14 | 559.58142892 | | 109677.96006793 | -0.887648 | | 1.179523 | 0.000145 | 0.064773 |
| 15 | 487.45749546 | | 109677.93647783 | -0.895138 | | 1.189480 | 0.000137 | 0.058542 |
| 16 | 428.42935810 | | 109677.91567386 | -0.901692 | | 1.198194 | 0.000130 | 0.053112 |
| 17 | 379.50829478 | | 109677.89719142 | -0.907475 | | 1.205882 | 0.000124 | 0.048364 |
| 18 | 338.51197736 | | 109677.88066302 | -0.912615 | | 1.212715 | 0.000118 | 0.044199 |
| 19 | 303.81680276 | | 109677.86579528 | -0.917214 | | 1.218830 | 0.000113 | 0.040529 |
| 20 | 274.19463088 | | 109677.85235040 | -0.921353 | | 1.224333 | 0.000109 | 0.037284 |

| n | $-E_{ns1/2}$ (cm$^{-1}$) [8] | A: $R_H(n)$ (cm$^{-1}$) | B: $R_H(n)$ (cm$^{-1}$) | $n_0/n-1$ | A: $\sqrt{CC_{nH}}$ | B: $\sqrt{CC_{nH}}$ | $\Delta\sqrt{CC_{nH}}$ | $\delta E_{nH}$ (MHz) |